# Non-additivities of the particle sizes hidden in model pair potentials and their effects on physical adsorptions

Ken-ichi Amano,[1,a)] Satoshi Furukawa,[2] Rina Ishii,[1] Ayane Tanase,[1] Masahiro Maebayashi,[1] Naoya Nishi,[2] and Tetsuo Sakka[2]

[1]Laboratory of Biophysical Chemistry, Department of Applied Biological Chemistry, Faculty of Agriculture, Meijo University, 1-501 Shiogamaguchi, Tempaku, Nagoya 468-8502, Japan

[2]Department of Energy and Hydrocarbon Chemistry, Graduate School of Engineering, Kyoto University, Kyoto 615-8510, Japan

[a)]Author to whom correspondence should be addressed: amanok@meijo-u.ac.jp

## Abstract

It is important to understand mechanism of colloidal particles assembly near a substrate for developments of batteries, heterogeneous catalysts, paints, and cosmetics. Knowledge of the mechanism is also important for crystallizations of the colloidal particles and proteins. In this study, we calculated the physical adsorption of colloidal particles on a flat wall by using the integral equation theory, wherein small and large colloidal particles were employed. In the calculation system, electric double layer potentials were used as the pair potentials. In some cases, it was found from the calculation results that the small particles are more easily adsorbed. The result is unusual from the viewpoint of the Asakura-Oosawa theory: we call it "reversal phenomenon". Then, we investigated mechanism of the reversal phenomenon. As a result, it was found that the inversion phenomenon originates from the non-additivities of the particle sizes. In addition, we invented the method to analyze the non-additivity in the pair potentials. The method will be useful for checks of various simulation results and developments of force fields for simulations of the colloidal particles and proteins.

# 1. Introduction

Many researches have been performed to understand the physical adsorption of colloidal particles on substrates. The deep understandings of the physical adsorptions are very important in various fields such as developments of drugs, cosmetics, paints, and batteries. The understandings will also lead to creation of new nanotechnologies. Therefore, it is important to capture deep understandings of the physical adsorptions.

The physical adsorptions have been actively studied by using integral equation theory,[1,2,3] density functional theory,[4,5] Monte Carlo[6,7] and molecular dynamics[8,9] simulations. In addition, it has been actively studied by experiments such as atomic force microscopy,[10] beam reflectivity,[11] and quartz crystal microbalance.[12] However, there are few research examples on the physical adsorption on a substrate focusing on non-additivity in terms of sizes.[5] Therefore, we started a research on this topic. The physical adsorption on the substrate is an important topic in development of designable coating technology, and is also important for crystallizations of colloidal particles[13] and proteins.[14]

In the present study, we used the integral equation theory[15,16] and model potentials (electric double layer potentials)[17,18] to calculate the contact density of particles in the vicinity of the substrate and investigate the adsorption properties. Two types of colloidal particles, large and small particles, were immersed in the system and their adsorption properties on a flat substrate were calculated using the integral equation theory. From the calculation, we obtained a result that the small particles are more adsorbed on the substrate than the large particles. This result is opposite to the result normally obtained from Asakura-Oosawa (AO) theory.[16,19,20] We call the phenomenon "reversal phenomenon". However, mechanism of the reversal phenomenon is unknown. Therefore, in this study, we challenge to elucidate mechanism of the reversal phenomenon.

For the elucidation, we focused on non-additive AO theory recently developed by our group.[21] As reported in the paper, the reversal phenomenon is also found from the non-additive AO theory. From the result, the non-additivity in terms of the size may be the cause of the reversal phenomenon. However, we do not know whether the model potentials used in the integral equation theory have non-additivity or not. Therefore, we try to obtain a method for checking existence of the non-additivity in the model potential.

The method of checking the existence of the non-additivity in the model potential is useful for the investigation of adsorption properties calculated from various simulations. This is because, for example, when a mysterious adsorption is obtained from a simulation, its mechanism can be considered from the viewpoint of the non-additivity. In addition, the method can be applied to improve shapes of the model potentials. Since the model potentials are used in various fields such as developments of drugs, cosmetics, paints, and batteries, the method we proposed has a significant impact on these fields.

## 2. Methods

### 2-1. Integral equation theory

We use the integral equation theory (statistical mechanics of simple liquids), which can calculate various correlation functions in a system. In the present study, Ornstein-Zernike (OZ) equation coupled with the hyper-netted-chain (HNC) equation[15,16] is applied. The simultaneous equation is named OZ-HNC. In the calculation system, two types of particles (1 and 2) and a flat wall are employed. Inputting the model pair potentials, the normalized density distributions of the particles 1 and 2 near the wall are obtained.

### 2-2. Model pair potentials

We apply the model pair potential derived from osmotic repulsive force between two substances. It is also called as electric double layer potential. That is, the model potential is Derjaguin-Landau-Verwey-Overbeek (DLVO) potential neglecting the van der Waals potential. The van der Waals force has a large effect at short distances, but little effect at longer distances. For this reason, we ignore the van der Waals potential and consider only the electric double layer potential. As the electric double layer potential, we used a pair potential obtained based on the linear superposition approximation (LSA).[17,18] The pair potential ($V_{ij}$) of the LSA model between particles $i$ and $j$ is expressed as

$$V_{ij}(r_{ij}) = \frac{4\pi\varepsilon_r\varepsilon_0 a_i a_j}{a_i + a_j}\psi_i\psi_j e^{\kappa(a_i+a_j)}e^{-\kappa r_{ij}}, \tag{1}$$

where $r_{ij}$ represents distance between the centers of the particles $i$ and $j$. $\varepsilon_r$, $\varepsilon_0$, $a_i$ ($i = 1$ or 2), $\psi_i$ ($i = 1$ or 2), and $\kappa$ are the relative permittivity of the solution, the vacuum permittivity, the core radius of particle $i$, the surface electric potential of particle $i$, and the reciprocal of Debye length, respectively. The pair potential of the LSA model between the wall and the particle $i$ is expressed as

$$V_{Wi}(h_{Wi}) = 4\pi\varepsilon_r\varepsilon_0 a_i \psi_W \psi_i e^{\kappa a_i} e^{-\kappa h_{Wi}}, \tag{2}$$

where $h_{Wi}$ represents distance between the surface of the wall and the center of the particle $i$.

### 2-3. Non-additive AO theory

The density distribution of particles near the wall can be calculated by using the AO theory. However, the calculation accuracy is poor when two bodies are not in contact. Then, in the present study, we calculate the density at the contact point (the closest distance between the particle and the wall). In addition, we insert the non-additivity in terms of sizes into the AO theory, which we call the non-additive AO theory. In the present system, there are two types of particles (1 and 2) and a flat wall whose surfaces are rigid. Distance between the centers of the particles $i$ (= 1 or 2) and $j$ (= 1 or 2) being $l_{ij}$ is described with non-additive parameter $\omega_{ij}$ as follows:[52223]

$$l_{ij} = \frac{1}{2}(d_i + d_j)(1 + \omega_{ij}), \tag{3}$$

where $d_i$ ($i = 1$ or 2) is diameter of the particle $i$ (see Fig. 1(a)). The value of $\omega_{ij}$ is larger than or equal to –1. When $\omega_{ij} = 0$, the particle pair distance is described by a general additive model. When $\omega_{ij}$ is smaller than 0, $l_{ij}$ is shorter than that of the additive model (and vice versa). Similarly, distance between the wall surface and the center of the particle $i$ (= 1 or 2) being $z_{Wi}$ is described with non-additive parameter $\omega_{Wi}$ as follows:

$$z_{Wi} = \frac{1}{2}(0 + d_i)(1 + \omega_{Wi}), \qquad (4)$$

where the subscript W represents the wall. The *zero* in Eq. (4) means that the right end of the line segment related to $z_{Wi}$ is at the wall surface (see Fig. 1(b)). When $\omega_{Wi} = 0$, the wall-particle pair distance is described by a general additive model. When $\omega_{Wi}$ is smaller than 0, $z_{Wi}$ is shorter than that of the additive model (and vice versa). In the calculation, pair of the wall and the paticle 1 is treated as the additive model (*i.e.*, $z_{W1} = d_1/2$).

Introducing the non-additivities explained above into the AO theory, the non-additive AO theory for the wall-particles system is obtained. The resultant equation is expressed as follows:[21]

$$g_{Wi,cp} = \exp\left[\sum_{j=1}^{n} \left\{\frac{2}{3}l_{ij}^3 - (z_{Wi} - z_{Wj})l_{ij}^2 + \frac{1}{3}(z_{Wi} - z_{Wj})^3\right\}\pi\rho_j\right], \qquad (5)$$

where $g_{Wi,cp}$ ($i$ = 1 or 2), $\pi$, and $\rho_j$ ($j$ = 1 or 2) are the normalized number density of the particle $i$ at the wall surface (contact point), the circular constant, and number density of the particle $j$ in the bulk, respectively. The subscript cp represents contact point.

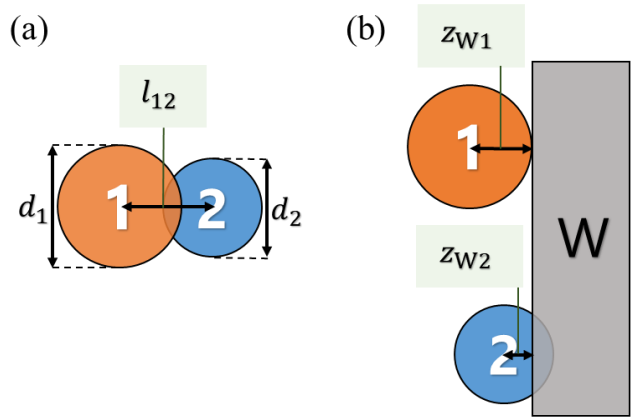

Figure 1. Schematics of contact distances $l_{12}$, $z_{W1}$, and $z_{W2}$.[21] In (a), the non-additive parameter $\omega_{12}$ is less than 0. In (b), the non-additive parameters $\omega_{W1}$ and $\omega_{W2}$ are 0 and less than 0, respectively.

## 3. Results and discussions

### 3-1. Density distribution calculated from the integral equation theory

We calculated the normalized number density of the particles 1 and 2 near the wall by using the integral equation theory. The LSA pair potentials (Eqs. (1) and (2)) were used to model the pair potentials in the system. The diameter of particle 1 was fixed at 12 nm, and the diameter of particle 2 was changed to 4 nm, 6 nm, 8 nm, and 10 nm. The volume fractions of particles 1 and 2 were both 5%, and the relative permittivity of water was 78.36 (25 °C).[24] The calculation results are shown in Fig. 2.

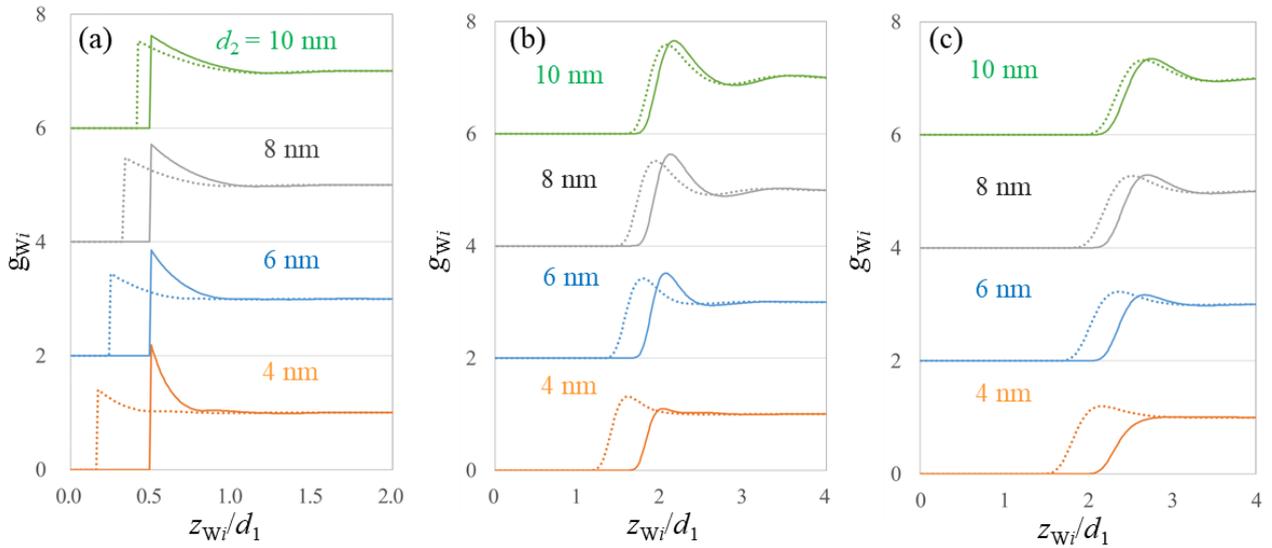

Figure 2. Normalized number density of particles near the wall calculated from the integral equation theory. The solid and dotted curves represent the normalized number densities of the particles 1 and 2, respectively. The volume fractions of the particles 1 and 2 are both constant (5%). All of the substances are rigid in (a). Debye length is 2 nm and all of the surface potentials are –30 mV in (b) and Debye length is 3 nm and all of the surface potentials are –20 mV in (c).

In Fig. 2(a), the system is all rigid one. In Fig. 2(b), Debye length is 2 nm and all of the surface potentials are –30 mV. In Fig. 2(c), Debye length is 3 nm and all of the surface potentials are –20 mV. The vertical and horizontal axes represent the normalized number density ($g_{Wi}$) of the particles and the normalized distance between the particle and the wall, respectively. In Fig. 2, the dotted curves colored in orange, blue, gray, and green are the normalized number densities of the particle 2 with the diameter 4 nm, 6 nm, 8 nm, and 10 nm, respectively.

In Fig. 2 (a), the contact density (peak density) of the particle 1 is always higher than that of the particle 2. The difference between the contact densities of the particles 1 and 2 becomes larger as the difference between the diameters is increased. The tendency is usually seen in the results from the AO theory. However, in Figs. 2(b) and 2(c), the contact density of the particle 1 is not always higher than that of the particle 2. That is, the reversal phenomenon is observed. As shown in Fig. 2(b), the contact density of the particle 2 (smaller one) with the diameter 4 nm is higher than that of the particle 1. Moreover, as shown in Fig. 2(c), the contact densities of the particle 2 with the diameters 4 nm and 6 nm are higher than those of the particle 1. From the viewpoint of the normal AO theory, this is mysterious phenomenon, mechanism of which will be explained in chapter 3-3.

The other trend in Figs. 2(a) and 2(b) to be considered is that the contact densities of the particles 1 and 2 decrease as the diameter of the particle 2 decreases regardless of the *constant* volume fraction, although it is difficult to see. The mechanism of the property will be also explained in chapter 3-3.

**3-2. Reversal of the contact density calculated from the non-additive AO theory**

From the integral equation theory, we found the reversal phenomenon. Such a kind of the phenomenon can be found from results of the non-additive AO theory. Hence, in this section, we show the reversal phenomenon obtained from the non-additive AO theory and compare the results from both theories.

The relationship between the contact density and each non-additive parameter can be investigated by using Eq. (5) and the results are shown in Fig. 3. In Figs. 3, $d_1$ = 12 nm, $d_2$ = 10 nm, $\rho_1$ = 5.53 × $10^{-5}$ nm$^{-3}$, $\rho_2$ = 9.55 × $10^{-5}$ nm$^{-3}$. The bulk number densities were set so that the volume fractions of the particles 1 and 2 are both 5%.

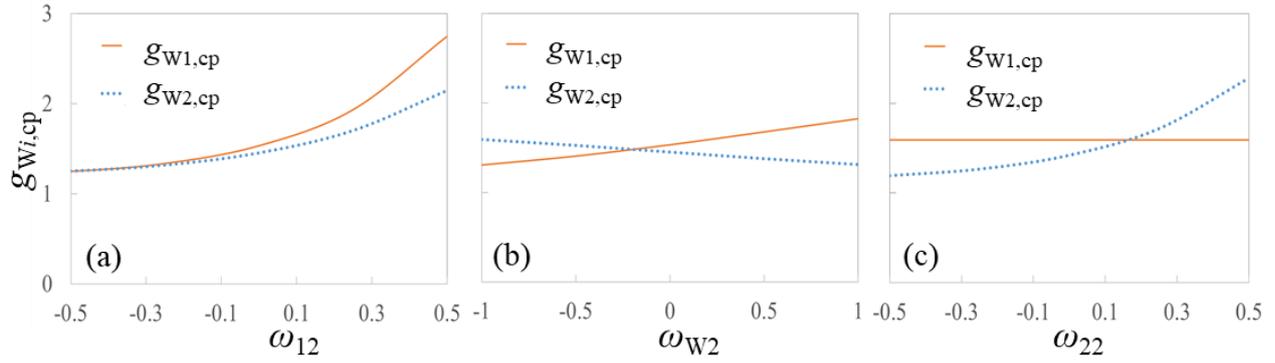

Figure 3. The contact densities of the particles 1 and 2 vs the non-additive parameters: (a) $g_{Wi,cp}$ vs $\omega_{12}$; (b) $g_{Wi,cp}$ vs $\omega_{W2}$; (c) $g_{Wi,cp}$ vs $\omega_{22}$. The volume fractions of the particles 1 and 2 are both constant (5%).

Fig. 3(a) shows relation between the contact density of the particle $i$ ($i$ = 1 or 2) and the non-additive parameter $\omega_{12}$. It was found that the contact density of both particles increased as $\omega_{12}$ increased despite the constant volume fractions. This is because, as $\omega_{12}$ increases, the bulk crowding increases, and hence, more particles tend to be adsorbed to reduce the crowding. The contact density of the particle 1 is significantly increased, because the particle 1 has the larger diameter, and thus the overlap volume between the particle 1 and the wall is larger than that between the particle 2 and the wall, leading to higher entropy gain.

Fig. 3(b) shows relation between the contact density of the particle $i$ ($i$ = 1 or 2) and the non-additive parameter $\omega_{W2}$. At $\omega_{W2}$ = 0, the contact density of particle 1 is higher than that of particle 2. The contact density of the particle 1 increases as $\omega_{W2}$ increases, but the contact density of the particle 2 decreases as $\omega_{W2}$ increases. Due to the different behaviors on the non-additive parameter, there is the reversal phenomenon. The reason of the behavior of the contact density of the particle 2 can be considered as follows. When $\omega_{W2}$ is negative, the particle 2 can penetrate into the wall more. It causes increases in the overlap volume between the particle 2 and the wall. Consequently, the adsorption of the particle 2 generates more entropy gain, which is the origin of the reversal phenomenon. We consider that the reversal phenomenon obtained from the non-additive AO theory is related to the reversal phenomenon from the integral equation theory, because the shapes of the repulsive areas of the LSA pair potentials also change depending on the LSA parameters. The detailed analyses of the shapes will be conducted in the next chapter. By the way, $\omega_{W2}$ can be controlled by manipulating pH, ionic strength, surface potential of the wall and the particles in the experimental system. Hence, the reversal phenomenon could be realized in the real system.

Fig. 3(c) shows relation between the contact density of the particle $i$ ($i = 1$ or $2$) and the non-additive parameter $\omega_{22}$. The contact density of the particle 1 is constant independent of $\omega_{22}$. The reason is that the excluded volumes of the particle 1 generated by the particles 1 and 2 are constant even when $\omega_{22}$ changes, and hence the overlap volume of the particle 1 and the wall at the contact does not change. On the other hand, the contact density of the particle 2 increases as $\omega_{22}$ increases. This is because, the excluded volume of the particle 2 generated by the same particle increases as $\omega_{22}$ increases, leading to the increase in the overlap volume of the particle 2 and the wall at the contact.

### 3-3. Non-additivities hidden in the LSA pair potentials

In chapter 3-1, we found the reversal phenomenon from the integral equation theory in which the smaller particle 2 was more adsorbed on the wall than the particle 1. In chapter 3-2, we explained the mechanism of the reversal phenomenon by showing the results from the non-additive AO theory. It was considered that the reversal phenomenon originates from the non-additive parameter $\omega_{W2}$. Then, in the present chapter, we investigate the non-additivities hidden in the model pair potentials used in the integral equation theory.

First, we focus on the LSA pair potential between the particles $i$ and $j$ (Eq. (1)). As shown in Eq. (1), the LSA pair potential does not have rigid surfaces, so the sizes of the particles $i$ and $j$ cannot be determined explicitly. Hence, we set $r_{ij,u}$ as the contact distance for the particles $i$ and $j$, where $r_{ij,u}$ represents the distance between the centers of the particles $i$ and $j$ when the pair potential is $u$ ($> 0$). From Eq. (1), $r_{ij,u}$ is expressed as follows:

$$r_{ij,u} = (a_i + a_j) - \frac{1}{\kappa} \ln \left\{ \frac{u(a_i + a_j)}{4\pi \varepsilon_r \varepsilon_0 a_i a_j \psi_i \psi_j} \right\}. \tag{6}$$

By the way, in the case of the rigid system in the non-additive AO theory, the contact distance between the particles $i$ and $j$ (= $l_{ij}$) was expressed in Eq. (3). In the case of the LSA pair potential, Eq. (3) can be rewritten with $r_{ij,u}$, $r_{ii,u}$, and $r_{jj,u}$ as follows:

$$r_{ij,u} = \frac{1}{2}(r_{ii,u} + r_{jj,u})(1 + \omega_{ij}). \tag{7}$$

Modifying the above equation, it is rewritten as

$$\omega_{ij} = \frac{2r_{ij,u} - (r_{ii,u} + r_{jj,u})}{r_{ii,u} + r_{jj,u}}. \tag{8}$$

Since the denominator $r_{ii,u} + r_{jj,u}$ in the above equation is always positive, $\omega_{ij}$ in the case of the LSA pair potential has the property that the sign (±) of $\omega_{ij}$ can be determined from the sign of the numerator $2r_{ij,u} - (r_{ii,u} + r_{jj,u})$. Substituting Eq. (6) into Eq. (8), the non-additive parameter can be obtained as follows:

$$\omega_{ij} = \frac{1}{\kappa(r_{ii,u} + r_{jj,u})} \ln\left\{\frac{4a_i a_j}{(a_i + a_j)^2}\right\}. \tag{9}$$

From Eq. (9), it is found that when $a_i = a_j$, $\omega_{ij} = 0$, and when $a_i \neq a_j$, $\omega_{ij} < 0$. This is because, $(a_i + a_j)^2 - 4a_i a_j = (a_i - a_j)^2 > 0$. The value of $\omega_{ij}$ deviates from zero as the difference between $a_i$ and $a_j$ increases. Interestingly, $\omega_{ij}$ in the case of the LSA pair potential does not take positive value, but negative value.

Second, we focus on the LSA pair potential between the wall and the particles $i$ (Eq. (2)). As shown in Eq. (2), the LSA pair potential does not have rigid surfaces, so the contact distance between the wall and the particle $i$ cannot be determined explicitly. Hence, we set $h_{Wi,u}$ as the contact distance for the wall and the particles $i$, where $h_{Wi,u}$ represents the distance between the surface of the wall and the center of the particle $i$ when the pair potential becomes $u$ (> 0). From Eq. (2), $h_{Wi,u}$ is expressed as follows:

$$h_{Wi,u} = a_i - \frac{1}{\kappa}\ln\left(\frac{u}{4\pi\varepsilon_r\varepsilon_0 a_i\psi_W\psi_i}\right). \tag{10}$$

By the way, in the case of the rigid system in the non-additive AO theory, the contact distance between the wall and the particles $i$ ($=z_{Wi}$) was expressed in Eq. (4). In the case of the LSA pair potential, Eq. (4) can be rewritten with $h_{Wi,u}$ and $r_{ii,u}$ as follows:

$$h_{Wi,u} = \frac{1}{2}(0 + r_{ii,u})(1 + \omega_{Wi}). \tag{11}$$

Modifying the above equation, it is rewritten as

$$\omega_{Wi} = \frac{2h_{Wi,u} - r_{ii,u}}{r_{ii,u}}. \tag{12}$$

Since the denominator $r_{ii,u}$ in the above equation is always positive, $\omega_{Wi}$ in the case of the LSA pair potential has the property that the sign ($\pm$) of $\omega_{Wi}$ can be determined from the sign of the numerator $2h_{Wi,u} - r_{ii,u}$. Substituting Eq. (10) into Eq. (12), the non-additive parameter can be obtained as follows:

$$\omega_{Wi} = \frac{1}{\kappa r_{ii,u}}\ln\left(\frac{8\pi\varepsilon_r\varepsilon_0 a_i\psi_W^2}{u}\right). \tag{13}$$

Since $u$ is an arbitrarily positive repulsive potential, it is difficult to judge the sign of the non-additive parameter $\omega_{Wi}$. Hence, we investigate which of the particles can be closer to the wall by comparing the numerators for the particles $i$ and $j$ in Eq. (12):

$$(2h_{Wi,u} - r_{ii,u}) - (2h_{Wj,u} - r_{jj,u}) = \frac{1}{\kappa}\ln\left(\frac{a_i}{a_j}\right). \tag{14}$$

When $a_i = a_j$, the value of the right-hand-side in Eq. (14) is zero. In this case, the non-additivities of the particles $i$ and $j$ in contact with the wall are the same. When $a_i \neq a_j$, the value of the right-hand-side in Eq. (14) is not zero. For instance, in the case of $a_i > a_j$, the value becomes positive. Hence, in this case, the particle $j$ can be closer to the wall than the particle $i$.

In summary, it has been found that the LSA pair potentials have non-additivities. The values of the non-additivities deviate from zero as the difference between the particle sizes increases. In chapter 3-1, the reversal phenomenon was found and it occurred when the difference between the sizes of the particles 1 and 2 is large (see Figs. 2(b) and 2(c)). In that situation, the particle 2 can be closer to the wall compared with the particle 1. The situation corresponds to $\omega_{W2} < \omega_{W1}$. Recalling the Fig. 3(b), when $\omega_{W2}$ is smaller, the contact density of the particle 2 has been larger than that of the particle 1. That is, the situations in both the integral equation theory (Figs. 2(b) and 2(c)) and the non-additive AO theory (Fig. 3(b)) are the same. Therefore, it is considered that the reversal phenomenon in Figs. 2(b) and 2(c) is caused by the relatively small value of $\omega_{W2}$. Although it is difficult to see, the other trend found in Figs. 2(a) and 2(b) has been that the contact densities of the particles 1 and 2 decrease as the diameter of the particle 2 decreases regardless of the *constant* volume fraction. This trend can be explained by using Fig. 3(a) and Eq. (9). As shown in Fig. 3(a), the both contact densities of the particles 1 and 2 decrease as $\omega_{12}$ decreases. From Eq. (9) it can be said that the $\omega_{12}$ decreases as the difference between the particle sizes is increased. Therefore, the contact densities of the particles 1 and 2 decrease as the diameter of the particle 2 decreases in Figs. 2(a) and 2(b).

### 3-4. Analyses of the non-additivities between particles

In chapter 3-3, we found the non-additivity in the LSA pair potentials. In the present chapter, we investigate the non-additivities in several model pair potentials between particles.

First, we investigate the Yukawa pair potential[25] proposed for colloidal particles in liquid. The Yukawa pair potential ($u_{ij}$) between the particles $i$ and $j$ is expressed as

$$u_{ij}(r_{ij}) = \frac{e^2}{4\pi\varepsilon_0\varepsilon_r}\left(\frac{z_i\exp(\kappa a_i)}{1+\kappa a_i}\right)\left(\frac{z_j\exp(\kappa a_j)}{1+\kappa a_j}\right)\frac{\exp(-\kappa r_{ij})}{r_{ij}}, \tag{15}$$

where $e$, $z_i$, $\kappa$, and $a_i$ are elementary charge, effective valence of the particle $i$, the reciprocal of Debye length, and core radius of the particle $i$. For analysis of the non-additivity, we set the distance $r_{ij,u}$ where $r_{ij,u}$ represents the distance between the centers of the particles $i$ and $j$ when the pair potential is $u$ ($> 0$). Using Eq. (15) and the Lambert W function, $r_{ij,u}$ can be expressed as follows:

$$r_{ij,u} = \frac{1}{\kappa}W\left[\frac{\kappa e^2}{u4\pi\varepsilon_0\varepsilon_r}\left(\frac{z_i\exp(\kappa a_i)}{1+\kappa a_i}\right)\left(\frac{z_j\exp(\kappa a_j)}{1+\kappa a_j}\right)\right]. \tag{16}$$

The non-additive parameter between the particles can be analyzed by using Eq. (8). Substituting Eq. (16) into Eq. (8), $\omega_{ij}$ for the Yukawa pair potential is given by

$$\omega_{ij} = \frac{2W\left[\frac{\kappa e^2}{u4\pi\varepsilon_0\varepsilon_r}\left(\frac{z_i\exp(\kappa a_i)}{1+\kappa a_i}\right)\left(\frac{z_j\exp(\kappa a_j)}{1+\kappa a_j}\right)\right]}{W\left[\frac{\kappa e^2}{u4\pi\varepsilon_0\varepsilon_r}\left(\frac{z_i\exp(\kappa a_i)}{1+\kappa a_i}\right)^2\right]+W\left[\frac{\kappa e^2}{u4\pi\varepsilon_0\varepsilon_r}\left(\frac{z_j\exp(\kappa a_j)}{1+\kappa a_j}\right)^2\right]} - 1. \tag{17}$$

To check the behavior of $\omega_{ij}$, we show the contour map of $\omega_{ij}$ (Fig. 4) where the horizontal and the vertical axes are size ratio (= $a_j/a_i$) and valence ratio (= $z_j/z_i$), respectively. In Fig. 4, the parameters are as follows: $z_i = 10$, $a_i = 10$ nm, $\kappa = 1/10^{-8}$ m$^{-1}$ = $10^8$ m$^{-1}$, $\varepsilon_r = 80$, $T = 300$ K, and $u = 10k_BT$. As shown in Fig. 4, the values of $\omega_{ij}$ are smaller than or equal to zero in the entire field in the map. For example, when the size and valence ratios are both 1, $\omega_{ij}$ is equal to zero. In addition, as the ratios depart from 1, $\omega_{ij}$ decreases. Hence, if a simulation is conducted by using the Yukawa pair potentials and the ratios are largely different from 1, the negative non-additivity arises, leading to decrease in the bulk crowding. As a consequence, the contact densities of the particles on the wall decrease.

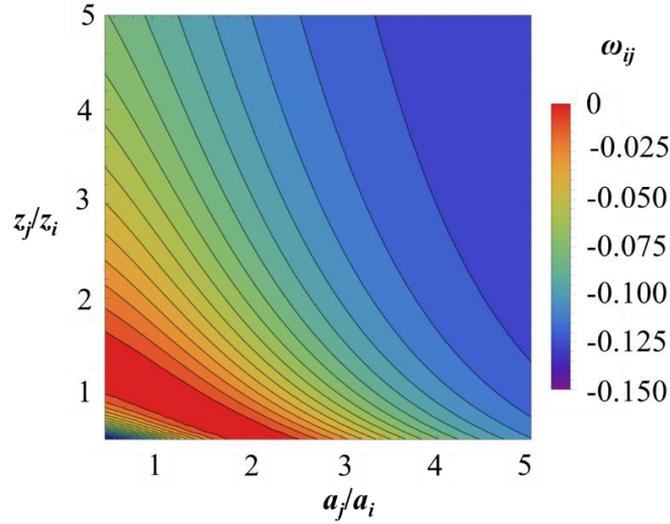

Figure 4. Non-additivity in the Yukawa pair potential between the particles.

Next, we investigate Lennard-Jones (LJ) potential. The LJ potential ($U_{ij}$) between the particles $i$ and $j$ is expressed as

$$U_{ij}(r_{ij}) = 4\varepsilon_{ij}\left[\left(\frac{\sigma_{ij}}{r_{ij}}\right)^p - \left(\frac{\sigma_{ij}}{r_{ij}}\right)^q\right], \tag{18}$$

where $\varepsilon_{ij}$ and $\sigma_{ij}$ are attractive parameter between the particles $i$ and $j$ and sum of the radii of the particles $i$ and $j$, respectively. Lorentz–Berthelot combining rules are applied in $\varepsilon_{ij}$ and $\sigma_{ij}$ as follows:

$$\varepsilon_{ij} = \sqrt{\varepsilon_{ii}\varepsilon_{jj}}, \tag{19}$$

$$\sigma_{ij} = \frac{\sigma_{ii} + \sigma_{jj}}{2}. \tag{20}$$

The indexes of $p$ and $q$ depend on the sizes of the particles. Here, we focus on the case that $p = 12$ and $q = 6$. In this case, the particles $i$ and $j$ are atomic-molecular sizes. For analysis of the non-additivity, we set the distance $r_{ij,U}$ where $r_{ij,U}$ represents the distance between the centers of the particles $i$ and $j$ when the pair potential is $U$ ($>$ 0). Using Eq. (18), $r_{ij,U}$ can be rewritten as

$$r_{ij,U} = \sigma_{ij}\left(\frac{1+\sqrt{1+U/\varepsilon_{ij}}}{2}\right)^{-1/6}. \tag{21}$$

Substituting Eq. (21) into Eq. (8), $\omega_{ij}$ for the LJ potential is given by

$$\omega_{ij} = \frac{(\sigma_{ii}+\sigma_{jj})\left(1+\sqrt{1+U/\sqrt{\varepsilon_{ii}\varepsilon_{jj}}}\right)^{-1/6}}{\sigma_{ii}\left(1+\sqrt{1+U/\varepsilon_{ii}}\right)^{-1/6}+\sigma_{jj}\left(1+\sqrt{1+U/\varepsilon_{jj}}\right)^{-1/6}} - 1. \tag{22}$$

To check the behavior of $\omega_{ij}$, we show the contour map of $\omega_{ij}$ (Fig. 5) where the horizontal and the vertical axes are size ratio (= $\sigma_{jj}/\sigma_{ii}$) and ratio of the attractive parameters (= $\varepsilon_{jj}/\varepsilon_{ii}$), respectively. In Fig. 5, the parameters are as follows: $\sigma_{ii}$ = 1 nm, $\varepsilon_{ii}$ = $10^{-21}$ J, $T$ = 300 K, and $U$ = $10k_{B}T$. There are positive and negative values of $\omega_{ij}$ in the map. For example, when the ratio of the attractive parameters is 1 (see the horizontal thick black line), $\omega_{ij}$ is always equal to zero. On the other hand, when the ratio of the size is 1, $\omega_{ij}$ is not always equal to zero, which can be recognized from the tilting thick gray line. If $\omega_{ij}$ is negative, crowding in the bulk is decreased. It unnaturally causes decrease in the contact density of the particles on the wall. Meanwhile if $\omega_{ij}$ is positive, crowding in the bulk is increased. It causes an increase in the contact density of the particles on the wall unnaturally.

From these results, it is considered that these behaviors should be concerned in simulations using model pair potentials (*e.g.*, the Yukawa pair potentials, the LJ potentials, etc.). We consider that the results in this study are useful for improvement of various model pair potentials.

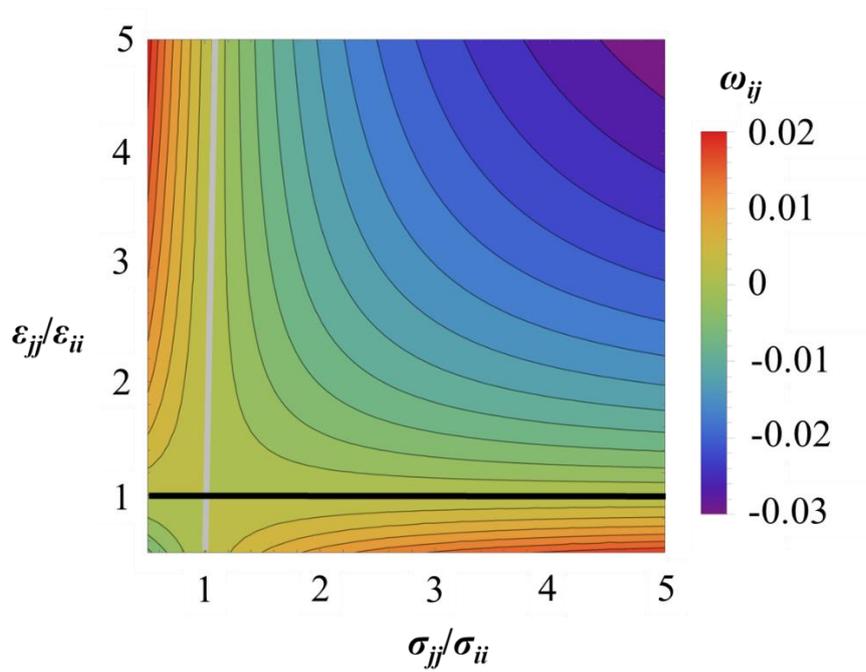

Figure 5. Non-additivity in the LJ potential between the particles. On the black horizontal line, $\omega_{ij}$ is always zero. Also on the gray line, $\omega_{ij}$ is always zero, but it is slightly tilting and the slope is positive.

## 4. Conclusions

We calculated the density distribution of the particles 1 (large) and 2 (small) by using the LSA pair potentials from the integral equation theory. From the calculation, we found that the contact density of the small particle is larger than that of the large particle. The result is abnormal from the viewpoint of the normal AO theory. Then, we focused on the non-additive AO theory, because the reversal phenomenon can be realized in that theory. Investigating the non-additivities in the LSA pair potentials, it was found that the non-additivities implicitly exist in the LSA pair potentials. It was also found that the non-additive parameters corresponding to the wall-particle pairs are the cause of the reversal phenomenon.

In the present study, we developed the method to investigate the existence of the non-additivity in a model pair potential. The method can be used for analysis of the contact density obtained from a simulation. For example, when a mysterious adsorption result is obtained, it is recommended to analyze the non-additivity by using the method we proposed. Since model potentials are widely used in various computational sciences such as developments of anti-cancer agents, vaccines, cosmetics, paints, and batteries, the method will also be widely used and support interpretations of the computational results. We believe that the method will also be applicable for improvements of shapes of the model pair potentials. Therefore, the present study will attract broad interests from wide computational research areas.

## Conflicts of interest

The authors declare no conflict of interest.

## Acknowledgements

This work was supported by JSPS KAKENHI Grant Number 20K05437 and partially supported by JSPS KAKENHI Grant Number 21K18935. We thank Hiroshi Onishi, Kota Hashimoto, and Hiroyuki Kato for supporting the research.